\documentclass[letters,twocolumn]{IEEEtran}
\IEEEoverridecommandlockouts
\usepackage{cite}
\usepackage{amsmath,amssymb,amsfonts}
\usepackage{algorithmic}
\usepackage{graphicx}
\usepackage{pgf,pgfplots,tikz,subfig}
\usepackage[export]{adjustbox}
\usepackage{url}
\usepackage{textcomp}
\usepackage{xcolor}
\usepackage[inkscapearea=page]{svg}
\usepackage{wrapfig}
\usepackage{caption}
\def\BibTeX{{\rm B\kern-.05em{\sc i\kern-.025em b}\kern-.08em
    T\kern-.1667em\lower.7ex\hbox{E}\kern-.125emX}}

\newcommand{\bx}{\textbf{\textit{x}}}
\newcommand{\bs}{\textbf{\textit{s}}}
\newcommand{\bz}{\textbf{\textit{z}}}
\newcommand{\by}{\textbf{\textit{y}}}

\begin{document}

\title{
An Efficient Wireless Channel Estimation Model \\
for Environment Sensing
}

\author{
\IEEEauthorblockN{Zainab R. Zaidi\IEEEauthorrefmark{1}, 
Tansu Alpcan\IEEEauthorrefmark{1},  
Christopher Leckie\IEEEauthorrefmark{2}, and 
Sarah Erfani\IEEEauthorrefmark{2}}\\
\IEEEauthorblockA{\IEEEauthorrefmark{1}
Department of Electrical and Electronic Engineering, The University of Melbourne, Australia \\
Email: $\{$zainab.raziazaidi, tansu.alpcan$\}$@unimelb.edu.au\\}
\IEEEauthorblockA{\IEEEauthorrefmark{2} School of Computing and Information Systems, The University of Melbourne, Australia \\
Email: $\{$caleckie, sarah.erfani$\}$@unimelb.edu.au}
\vspace{-12mm}}

\maketitle

\begin{abstract}
This paper presents a novel and efficient wireless channel estimation scheme based on a tapped delay line (TDL) model of wireless signal propagation, where a data-driven machine learning approach is used to estimate the path delays and gains. The key motivation for our novel channel estimation model is to gain environment awareness, i.e., detecting changes in path delays and gains related to interesting objects and events in the field. The estimated channel state provides a more detailed measure to sense the field than the single-tap channel state indicator (CSI) in current OFDM systems. Advantages of this approach also include low computation time and training data requirements, making it suitable for environment awareness applications.

We evaluate this model's performance using Matlab's ray-tracing tool under static and dynamic conditions for increased realism instead of the standard evaluation approaches that rely on classical statistical channel models. Our results show that our TDL-based model can accurately estimate the path delays and associated gains for a broad-range of locations and operating conditions. Root-mean-square estimation error was less than $10^{-4}$, or $-40$dB, for SNR $\geq 60$dB in all of our experiments. Our results show that interference of a flying drone on signal multipaths, in a preliminary experiment, can be detected in estimated channel states which, otherwise, remains obscured in conventional CSI.
\end{abstract}

\begin{IEEEkeywords}
Channel estimation, tapped delay line model, ray-tracing, machine learning, environment awareness
\end{IEEEkeywords}
\vspace{-5mm}
\section{Introduction}

Wireless environment sensing is a cutting-edge approach that enhances the surveillance capacity of conventional systems, e.g., sensors, without the need of additional hardware. By continuously monitoring wireless signals within and around critical facilities and infrastructure, unusual or unauthorized activities, security breaches or tampering with equipment can be identified \cite{Moghaddam2023,Wang2017,Mao2019,Rai2021}. A number of existing methods use Channel State Indicators (CSI) \cite{Wang2017,Mao2019} calculated from pilot signals to detect changes in the environment. However, CSI estimation in a standard communication system is not designed to detect subtle environmental changes. Rather, it represents the cumulative impact of the surroundings on the signal and could result in erroneous sensing performance. Moreover, the background noise and the time-varying dynamics of wireless channels remain major challenges to sensing efforts. 

Recently, data-driven machine learning techniques for estimating complex and non-linear system dynamics have led to the development of novel wireless channel estimation (CE) and management approaches \cite{He2018,OShea2019,Ye2020}. However, machine learning (ML)-based channel estimation models require substantial computation time and data for training, which makes them unsuitable for real-time applications. Moreover, they are generally black-box, or non-interpretable, and the complex patterns learned by the model cannot be used to derive key insights about the radio environment. 

In this paper, we propose a ``gray-box" approach, where we use a Tapped-Delay Line (TDL) model of the wireless multipath channel and learn its parameters through a ML-inspired data-driven approach. Our results show that this approach not only provides better channel estimation than conventional methods, such as Least Squares, but the estimated channel state is also a more detailed measure to sense the field in an environment awareness application. The detection of anomalies from sensed data requires further development and will be discussed in future work. This paper is focused on the design and performance of our novel CE method but, as a proof of concept, we include a preliminary case study in our results section of unmanned air vehicle (UAV or drone) detection using our channel estimation model with conventional wireless signals. 

Prior efforts in regards to TDL/multipath channel estimation model, include ITU's (International Telecommunication Union) recommendations\footnote{\url{https://www.itu.int/rec/R-REC-P.1407/en}} for parameterization of multipath environments, which use power thresholds to determine the number of multipath components. Not meant for real-time CE, ITU recommendations are for characterization of different scenarios. Some efforts using ML have also been proposed, e.g., \cite{Jaradat2021} models the tap and gain estimation problem as multi-class classification using a deep neural network and requires training of the model by a dictionary of channel models. This work improves the CE performance of a previous method \cite{Cheng2018}. which formulates the estimation of taps for specific transmitted and received signals in a Multi-Input Multi-Output (MIMO) system, as an optimization problem. 

Moreover, the existing CE approaches, in general, use statistical approaches, such as, Rayleigh or Rician fading models, to evaluate performance, which are inadequate for environment sensing experiments. In this paper, we use MATLAB's ray-tracing tool for increased realism to evaluate our CE method and subsequent sensing experiments. Our CE approach leverages the respective strengths of classical and ML techniques and efficiently calculates channel state in sufficient detail to facilitate environment sensing. In particular, our contributions in the paper are as follows:
\begin{itemize}
    \item We propose a novel and efficient data-driven channel estimation approach based on a neural network with one hidden layer using the pilot/reference signals already available in wireless communication systems. 
    \item We evaluate the CE model using Matlab's ray-tracing tool instead of the classical approach of employing statistical channel models, for increased realism.
    \item Our CE model shows accurate estimation (root-mean-square error below $10^{-4}$ or $-40$dB) for different static and dynamic scenarios with reasonable SNR ($\geq 60$dB).
    \item We also show the potential of our CE model in finding outlier events in the neighbourhood of the transmitter and receiver via a drone detection scenario.
\end{itemize}

\section{System Model}\label{sec:model}

The signal from a wireless transmitter can be represented as \cite{goldsmith_2005}:\vspace{-4mm}
\begin{eqnarray*}
    x(t) = \operatorname{Re}\{s(t)e^{j2\pi f_ct}\},
\end{eqnarray*}
where $s(t)$ is the baseband signal and $f_c$ is the carrier frequency. The discrete time versions of the baseband and transmitted signal are $\bs_n$ and $\bx_n$ respectively, both vectors of length $K$. Considering that the signal will reach the receiver through line-of-sight and various reflected paths, which experience different Doppler shifts, the received signal $y(t)$ can be written as \cite{goldsmith_2005}:
\begin{eqnarray*}
    y(t)\hspace{-0.5mm}=\hspace{-0.5mm}\operatorname{Re}\left \{ \hspace{-1.5mm} \sum_{i=0}^{D(t)-1} \hspace{-3mm}{c_i(t)s(t\hspace{-1mm}-\hspace{-1mm}\tau_i(t))e^{j(2\pi f_c\hspace{-0.5mm}(t-\tau_i(t)) + {\phi}_i(t))}\hspace{-1mm}+\hspace{-1mm}w(t)}\hspace{-1mm} \right\},
\end{eqnarray*}
where $c_i(t)$, $\tau_i(t)$, and ${\phi}_i(t)$ are the gain, delay, and Doppler shift of the $i$th path, $i=0$ corresponds to the line-of-sight and there are $D(t)$ distinct multipaths. $w(t)$ is the white Gaussian noise. In a time-varying channel model, the parameters $\{c_i(t), \tau_i(t), {\phi}_i(t)\} ~\forall{i}$, and $D(t)$ are random processes. The channel impulse response can be written as:
\begin{eqnarray}
    h(\theta_h(t)) = \sum_{i=0}^{D(t)-1}{c_i(t)\delta(t-\tau_i(t))e^{j{\phi}_i(t)}},
\end{eqnarray}
where $\theta_h(t)$, the parameters to describe channel impulse response, are defined as:
\begin{eqnarray}
    \theta_h(t) =  \{c_i(t), \tau_i(t), {\phi}_i(t) \},~i = 0, \cdots, D(t).
\end{eqnarray}

As can be seen in the above expression, the total phase shift experienced by the signal is $-2\pi f_c\hspace{-0.5mm}\tau_i(t) + {\phi}_i(t)$ due to multipaths and Doppler effect. The discrete time received signal and received baseband signal are denoted by $\by_n$ and $\bz_n$ respectively, which are also considered as $K$ length vectors. 

\section{TDL-NN Channel Estimation Approach}\label{sec:CE}

\begin{figure}[tbh!]
\begin{center}
\includegraphics[scale=0.45,clip = true]{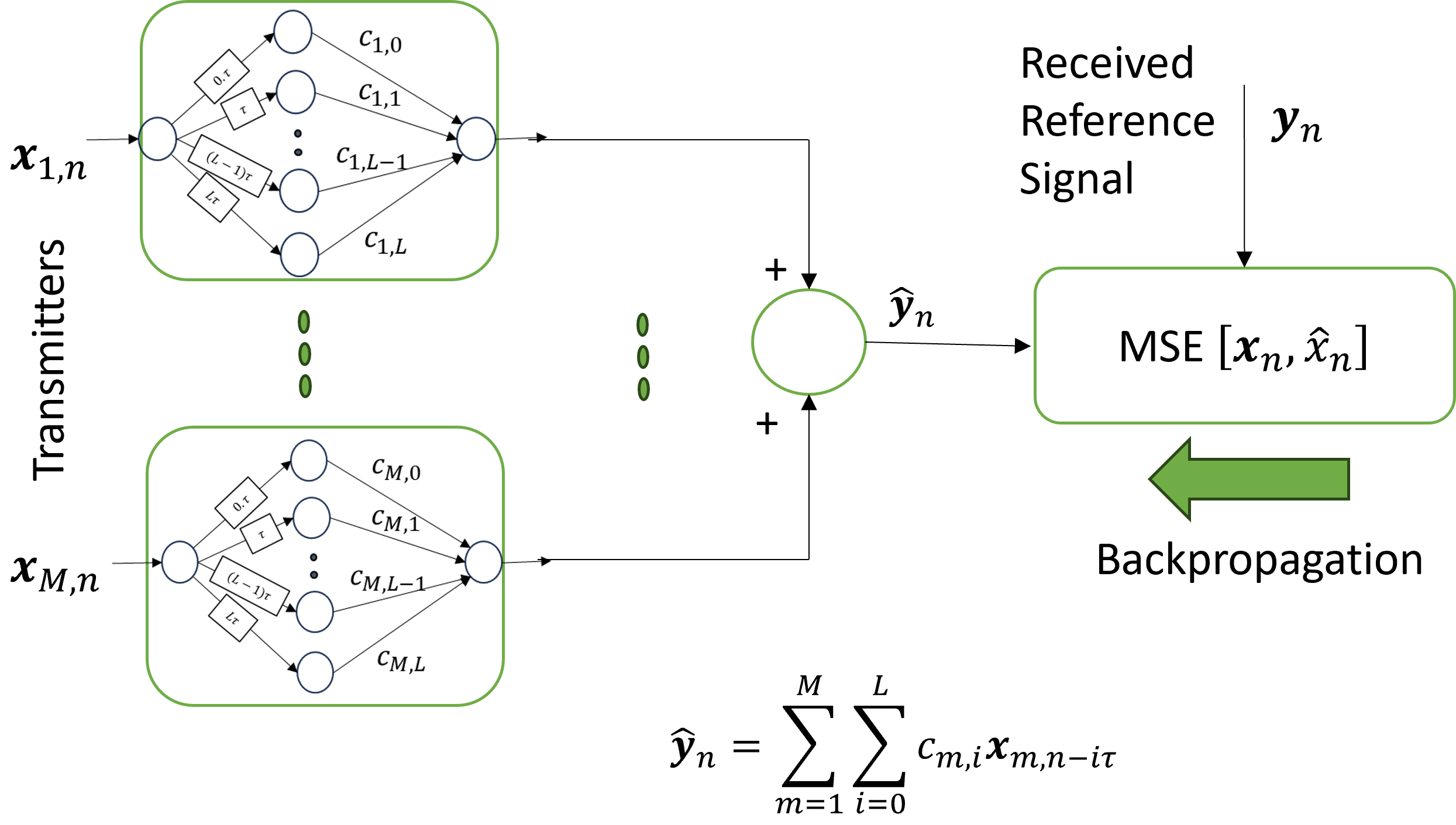}\vspace{-2mm}
\caption{TDL-NN channel estimation for a scenario with multiple transmitters and one receiver.}
\label{fig:method}
\vspace{-5mm}
\end{center}
\end{figure}

Figure \ref{fig:method} shows our approach for estimating the path delays and gains. The basic unit of our model, called TDL-NN, is a neural network (NN) with a single input and single output and one hidden layer representing $L$ taps of constant delays in multiples of $\tau$. The parameters $L$ and $\tau$ should cover all possible distinct path delays. 

All wireless communication systems use pilots or reference signals, which are known sequences of symbols transmitted over the wireless channel, and the respective received signals are used for channel state estimation. Figure \ref{fig:method} shows $\bx_{m,n}$ transmitted reference signals from one primary transmitter and $M-1$ multiple interferers. The received reference signal is $\by_{n}$. Our CE model is trained using $\bx_{m,n}$ and $\by_n$ for $m = {1,~\cdots,~M}$ and $n = {1,~\cdots,~N}$, where $N$ is the available reference symbols for training. The model shifts each input or transmitted signal, $\bx_{m,n}$, from $0 \times \tau$ to $L \times \tau$, where $\tau$ is the tap delay resolution in the hidden layer. It then computes the weights of the output layers, or coefficients $c_{m,i}$ for all taps and also for all $M$ sub-models, by minimizing the MSE (Mean Square Error) loss function, i.e., 
\[\frac{1}{N}\sum_{n=1}^N  \|\by_n - \hat{\by}_n\|^2_2,\] where $\hat{\by}_n = \sum_{m=1}^M\sum_{i=0}^Lc_{m,i}\bx_{m,n-i\tau}$.

Our implementation in PyTorch uses $N$ as the block-size and a learning rate of $0.01$. The computation complexity of our model grows as $\mathcal{O}(N \times L \times e)$, where $e$ is the number of epochs. As an option, we also implemented weight pruning in our model to push the smaller weights to 0. Note that our model uses the time shift idea as in conventional time delay neural networks \cite{Hassoun1995}, which were proposed for time series analysis, but otherwise it is a customized model to suit our application of channel estimation.

Path delays and gains change with time as mobile users or objects in the environment move. Weather conditions, such as, rain and wind also impact path gain. For a dynamic environment, the multipath delays and gains should be updated as new pilot symbols are received. Figure \ref{fig:method} shows a snapshot of our approach, which estimates $c_{m,i}(n)$ for each path tap $i$ at time $n$ by updating the model weights using pilot transmitted and received signals $\bx_{m,n}$ and $\by_{n}$. Note that we are not explicitly estimating Doppler phase shift $\phi_{i}(t)$ associated with the path $i$ from the model, instead we are tracking the variations in path delays and gains.

% Wrong section but places the figure at the top of the page
\begin{figure}[tph!] 
  \begin{minipage}[t]{0.6\textwidth}\vspace{-5mm}
    \subfloat[]{
        \includegraphics[scale=0.2,valign=c]{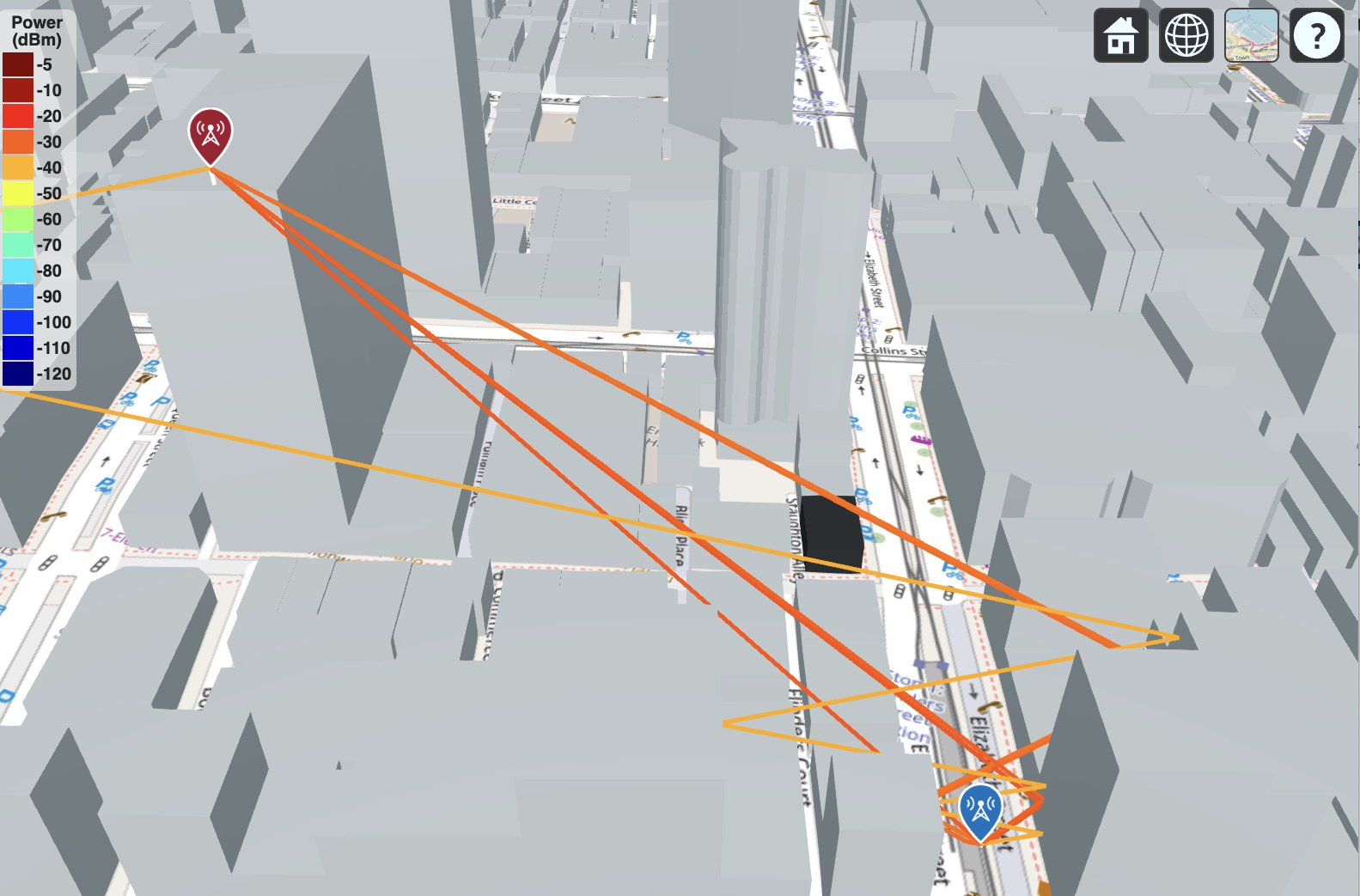}
        \label{fig:raytrace-static-1pair-scene}}
    \subfloat[]{
        \includegraphics[scale=0.15,valign=c]{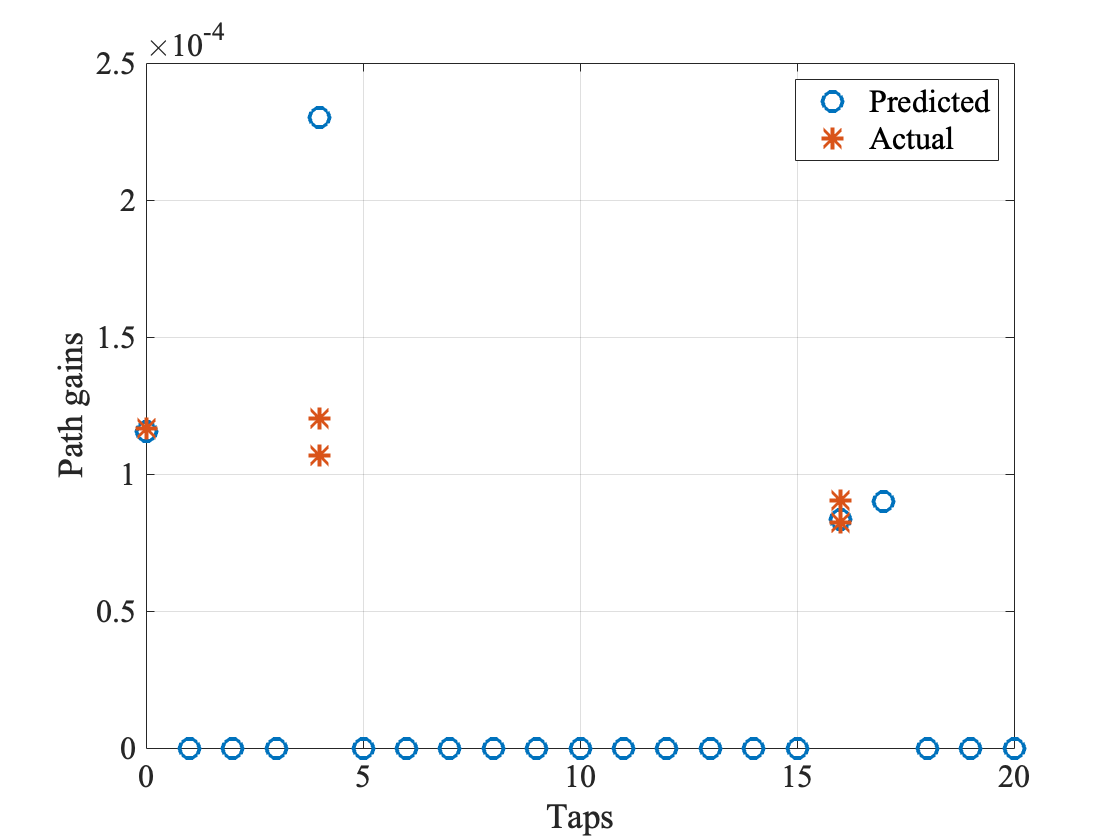}
        \label{fig:raytrace-static-1pair-results}} \vspace{-3mm}
        \\ 
    \subfloat[]{
        \includegraphics[scale=0.2,valign=c]{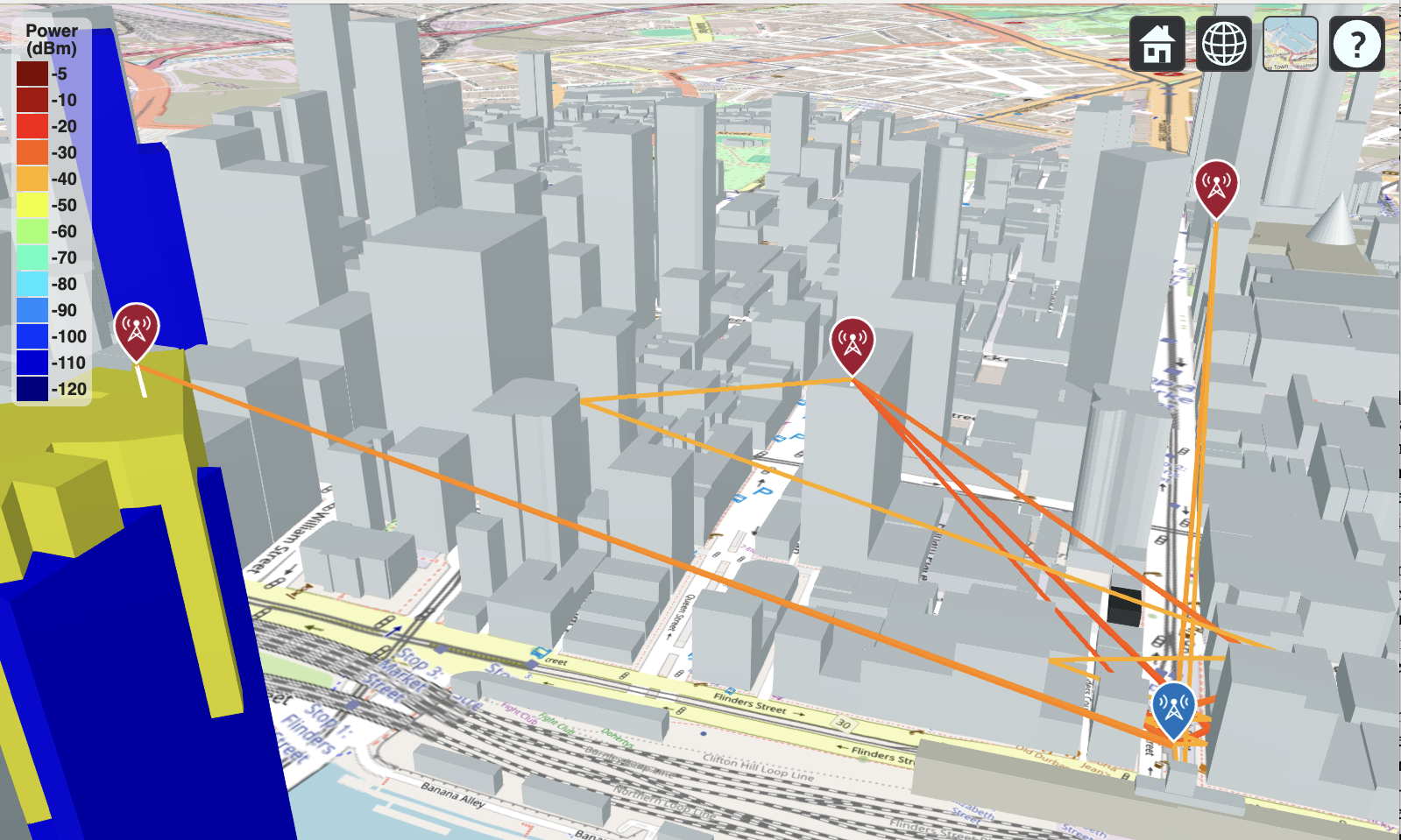}
        \label{fig:3tx-scene}}
    \subfloat[]{
         \includegraphics[scale=0.15,valign=c]{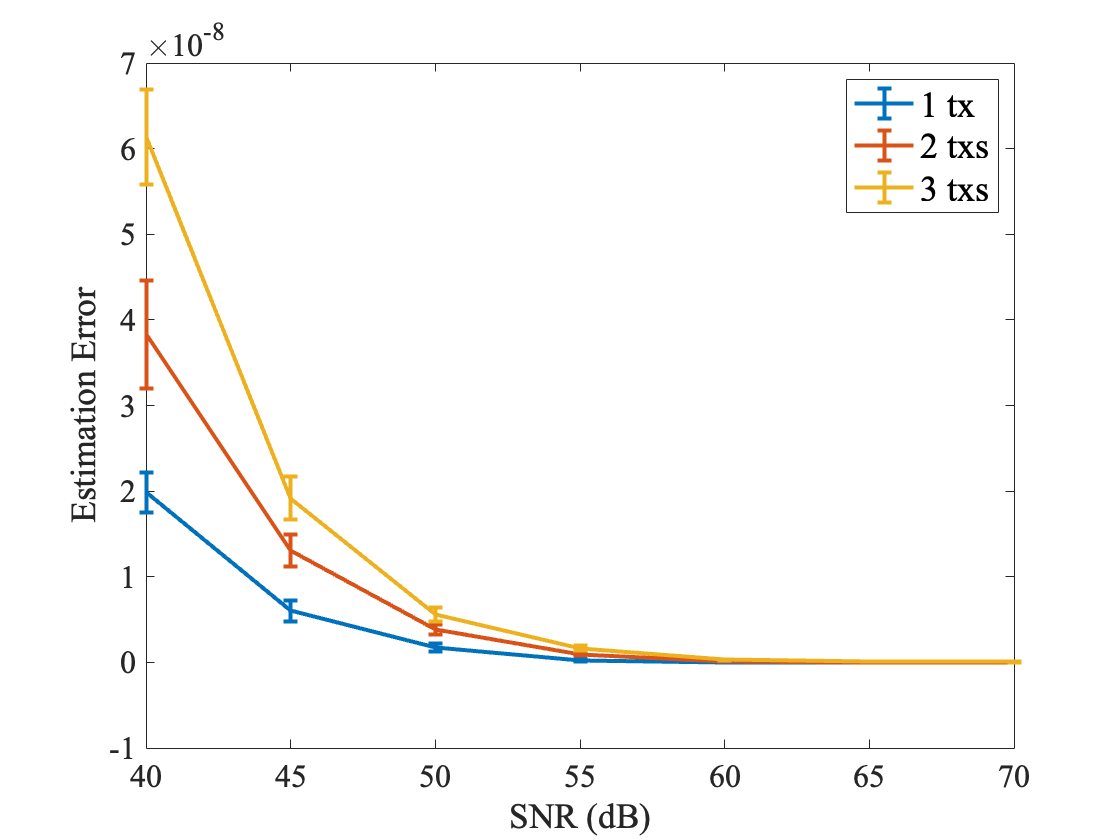}
        \label{fig:snr-error}}\vspace{-2mm}
    \caption{Static multipath scenario, (a) scene with one transmitter and receiver, (b) actual and estimated path delays and gains, tap 4 has 2 different rays and the estimation shows the combined effect, (c) scene of (a) with two interferers, (d) estimation error versus SNR for scenarios of one transmitter with no interferer (1 tx), one interferer (2 txs), and 2 interferers (3 txs).
  }
    \vspace{-5mm}
 \end{minipage}
\end{figure}

Moreover, we remark that our TDL neural network is a realization of a typical Steepest Gradient Descent (SGD) algorithm iteratively computing coefficients of a linear predictor when the objective function is to minimize mean square error between actual observations and predictions. However, using SGD in a neural network setting with parallelization of computations is much more effective and an order of magnitude faster than the conventional iterative approach.

%Furthermore, in systems using OFDM modulation, such as 4/5G or LTE, reference signals or pilots are embedded in subframes along with the payload. Channel estimation requires these pilots to be extracted first and then used to estimate the channel. In this case, baseband signals can be used for CE instead of transmitted or received signals with a carrier. 

\section{Experimental Results}\label{sec:experiments}

\subsection{Ray-tracing Simulations}

In order to validate our TDL-based CE approach, we used Matlab's ray-tracing tool to generate received signal data. To the best of our knowledge, this is the first CE paper that uses ray-tracing data for validation. Prior approaches have used statistical channel models, such as Rayleigh or Rician models, to generate the data \cite{OShea2019,Ye2020}. The use of ray-tracing not only allows us to assess the performance of our CE method by comparing it with the ground truth, it also facilitates testing the model efficacy under a wide-range of practical scenarios.

\subsubsection{Static Multipath Scenario}

We placed a transmitter and receiver in the Melbourne Central Business District scenario as shown in Fig.\ \ref{fig:raytrace-static-1pair-scene}. The transmitter is configured to transmit a QPSK modulated signal with $900$MHz carrier frequency and $50$dBm power. It is placed at a height of 7m on the roof of a tall building and uses a directional antenna with azimuth angle of 30\textdegree. The receiver antenna is at a height of 1m from the ground with an angle of 120\textdegree. 

The maximum number of reflections in Matlab's ray-tracing tool can be set to 10 (R2023b). White Gaussian noise is added to the faded signal with SNR of $80$dB. The signal is amplified by $30$dB on reception. $10{,}000$ symbols, with $100$ samples per symbol, are generated and transmitted through the multipath channel. Each sample duration is $20$ns corresponding to a sampling frequency of $5 \times 10^7$ samples/second. Unless otherwise stated, all the experiments in this paper uses the same configurations.

\newpage
\begin{wrapfigure}{L}{0.12\textwidth}
\rule{0pt}{11.5cm}% height of wrapfigure
\end{wrapfigure}

As shown in Fig.\ \ref{fig:raytrace-static-1pair-results}, our TDL-NN model accurately finds channel taps and gains. Tap 4 shows some discrepancy, but the estimated gain is the collective impact of two rays with the same path delay. %With weight pruning, smaller tap coefficients can be pushed to 0. However, for an unknown or time varying channel, this may not be advisable. 
We also calculated the RMS (Root Mean Square) estimation error for different SNR conditions, as shown in Fig.\ \ref{fig:snr-error}. Here, each point is averaged over 10 different estimation runs with $10{,}000$ training samples. Estimation error remains small for low noise conditions. %, e.g., at SNR=10dB, all estimated tap gains are similar, and we are not even able to detect the correct taps corresponding to the path delays experienced by the received signal. 

We added interferers in our example as shown in Fig.\ \ref{fig:3tx-scene} and estimated path delays and gains for all transmitted signals (Fig.\ \ref{fig:snr-error}). As expected, the presence of multiple transmitters worsens the estimation error. However, the estimation accuracy of TDL-NN model remains high, for reasonable SNR, despite the presence of in-band interference.

Table \ref{tab:ms_error} shows the estimation error vs.\ SNR for two urban scenarios (Melbourne Central Business District, Victoria, Australia and Manhattan, NY, USA), two suburban scenarios (Mount Waverley, Victoria, Australia and Chapelford, UK), and two rural areas (Wodonga, Victoria, Australia and Golden, CO, USA). The urban scenarios are more challenging due to the number of multipaths, whereas suburban areas have comparatively smaller number of reflections, but line-of-sight may be obstructed. Also, path attenuation may be excessive due to larger distances between users and base-stations in a sparse coverage situation. Rural areas on the other hand have mostly just the line-of-sight. Our TDL based CE approach performs exceptionally well for all conditions for reasonable SNR conditions as shown in Table \ref{tab:ms_error}.% shows the utility of our approach is a variety of circumstances and settings.

\begin{table}[!tbh]
    \centering
    \captionsetup{justification=centering}
    \caption{RMS Estimation error versus SNR \newline Urban, suburban, and rural scenarios}
    \begin{tabular}{l|c|c|c}\hline\hline
    Scenario & SNR = 40dB & SNR = 60dB & SNR = 80dB\\
    \hline
       Urban, Melbourne  &  $2\times10^{-8}$ & $3\times10^{-11}$ & $2\times10^{-15}$ \\
       Utban, NY & $2\times10^{-8}$ & $1\times10^{-11}$ & $2\times10^{-14}$ \\
       Suburban, VIC & $2\times10^{-8}$ & $6\times10^{-12}$ & $1\times10^{-14}$ \\
       Suburban, UK & $2\times10^{-8}$ & $1\times10^{-11}$ & $1\times10^{-14}$ \\
       Rural, VIC & $2\times10^{-8}$ & $2\times10^{-10}$ & $9\times10^{-14}$ \\
       Rural, CO & $2\times10^{-8}$ & $2\times10^{-10}$ & $1\times10^{-15}$ \\
       \hline
       \hline
    \end{tabular}
    \label{tab:ms_error}
\end{table}
\vspace{-1mm}

%\vspace{-3mm}
%\begin{figure}[tbh!] 
%\centering
%  \includegraphics[scale=0.165]{multi-scenario6.png}
%  \caption{Estimation error versus SNR of TDL-NN channel estimation model in different urban, suburban, and rural scenarios.}
%  \label{fig:summary-stat}
%    \vspace{-2mm}
%\end{figure}

\subsubsection{Mobile Receiver Scenario}

\begin{figure*}[tph!] 
    \subfloat[]{
        \includegraphics[width= 0.33\textwidth,scale=0.12]{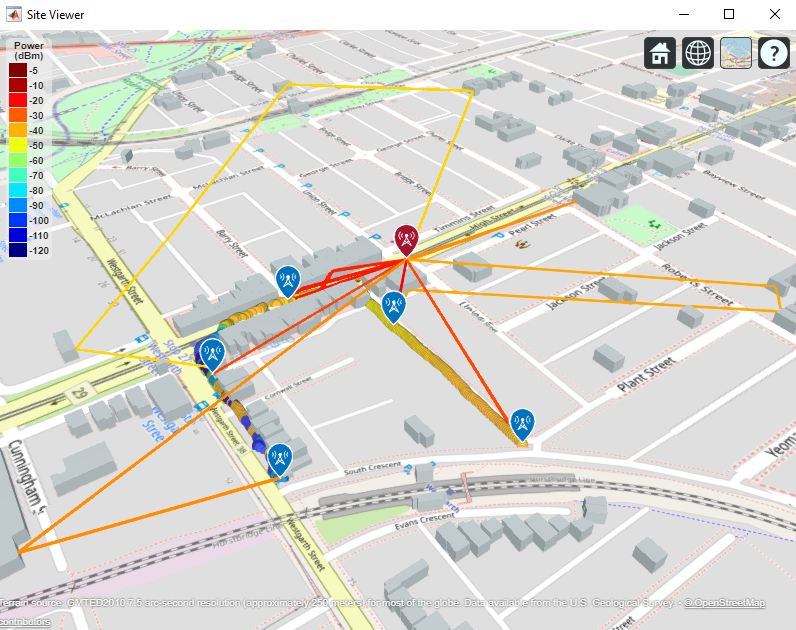}
        \label{fig:moving_scene}}
    %\hfill
    \subfloat[]{
        \includegraphics[width= 0.33\textwidth,scale=0.18]{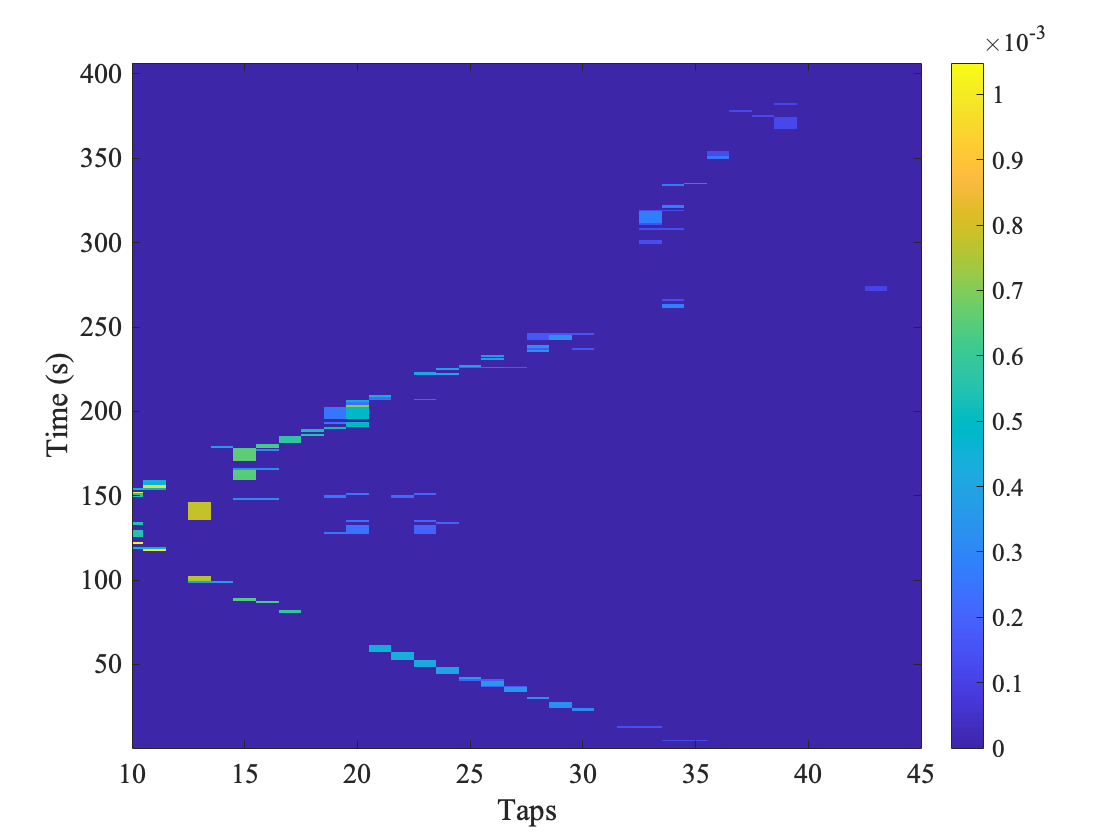}
        \label{fig:moving_true}}\hspace{-2mm}
    %\hfill
    \subfloat[]{
        \includegraphics[width= 0.33\textwidth,scale=0.18]{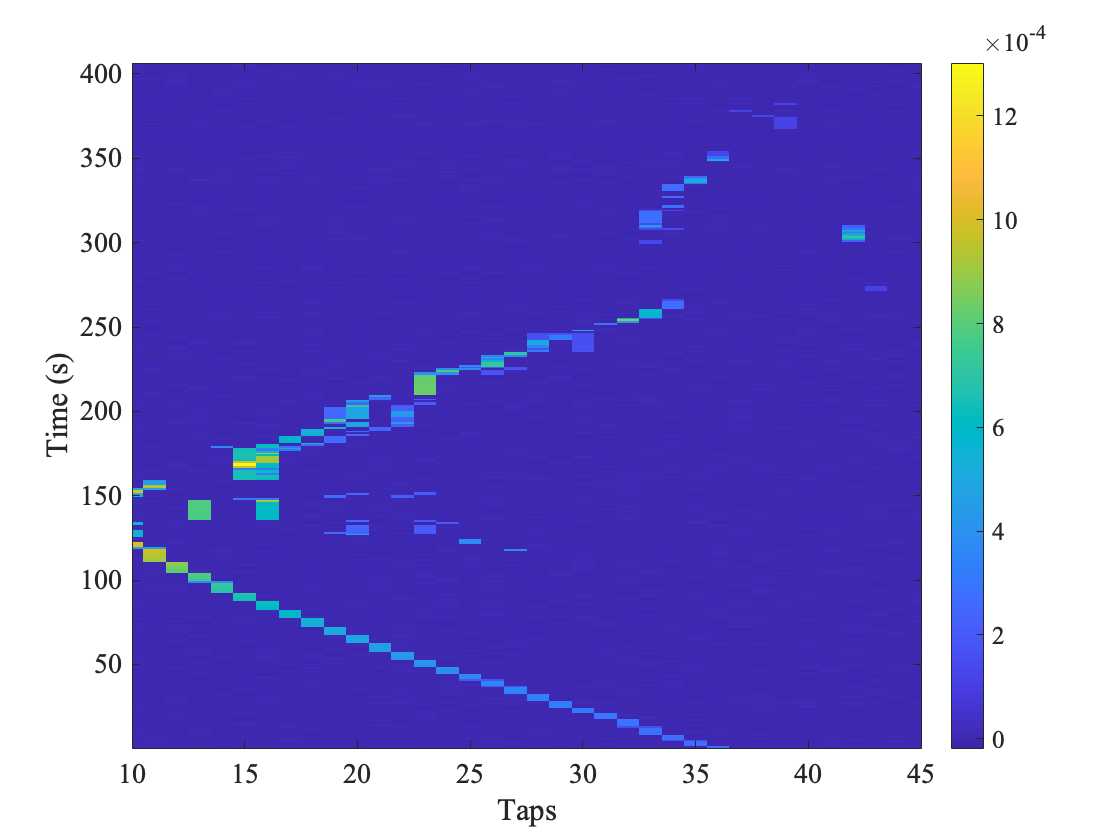}
        \label{fig:moving_results}}
    \caption{Changes in signal multipaths as the mobile user moves around a city block, (a) scene with changing rays when the receiver is at different locations, and (b) actual and (c) estimated taps for each observation.}
    \vspace{-5mm}
\end{figure*}
In this experiment, we used an open-access %\footnote{\url{https://www.openstreetmap.org/user/melbournefan/traces/3521343}} 
GPS track of a walker around a block of streets, representing a receiver, as shown in Fig.\ \ref{fig:moving_scene}. Here a transmitter is placed 10m above the roof of a nearby building. Other configurations are the same as the static scenario. The colours of the GPS track show the signal strength at that location, Figure \ref{fig:moving_scene} shows the receiver at multiple locations during the walk and changing multipath conditions. 

This track has 405 points, sampling the walker's location every second. We generated $1000$ transmitted symbols for each location, each symbol is sampled 150 times at the rate of $5 \times 10^7$ samples/second. The estimation results are shown in Fig.\ \ref{fig:moving_results} for SNR of $80$dB, where it can be seen that the TDL-NN model is tracking the changing multipaths at every step accurately. %The discrepancy in actual and estimated path gains is due to multiple paths with the same delay, and estimated gains are showing the cumulative impact. The lower left-hand image in Fig.\ \ref{fig:moving_results} shows some 
Sporadic path delays observed during the walk and respective estimations are shown by lighter pixels.% and the image of estimated tap gains shows that our model was able to catch them as well.

% Second moving experiment 
%lunch bike ride https://www.openstreetmap.org/user/wilfisher/traces/11148874
% Table will be added Estimation error (mean, std) vs. SNR 40, 30, 20 dB

We also repeated the experiment with SNR of $60$dB and $40$dB and also for a GPS trace, containing 400 points, of a bike ride in NY, USA. The RMS estimation error for both scenarios and for different SNR conditions are shown in Table \ref{tab:mobile_error}, where it can be seen that the TDL-NN model accurately estimates the dynamic channels.

\begin{table}[!tbh]
    \centering
    \captionsetup{justification=centering}
    \caption{RMS Estimation error versus SNR \newline Mobile receiver scenarios} 
    \begin{tabular}{l|c|c|c}\hline\hline
    Scenario & SNR = 40dB & SNR = 60dB & SNR = 80dB\\
    \hline
       Walk, Melbourne  &  $2.4\times10^{-4}$ & $3.4\times10^{-5}$ & $2.4\times10^{-5}$ \\
       Bike-ride, NY & $2.4\times10^{-4}$ & $2.5\times10^{-5}$ & $3.7\times10^{-6}$ \\
       \hline
       \hline
    \end{tabular}
    \label{tab:mobile_error}
\vspace{-5mm}
\end{table}

\subsection{BER Comparison with other Channel Estimation Methods}
As discussed above, our TDL-NN model is a novel approach specifically designed for environment sensing, as classical channel estimation methods, in general, only provide one-tap CSI. We compare the sensing performance of our channel estimates with classical CSI in the next section. In order to compare our method with prior art, we calculate the BER (Bit Error Rate) after equalization and demodulation using the channel estimates provided by our TDL-NN model and compare it with the BER from other CE methods. The results are given in Table \ref{tab:ber}.

%\vspace{-2mm}
%\begin{figure}[tbh!] 
%\begin{center}
%\includegraphics[scale=0.168]{comparative-study5_tmp.png}
%  \caption{Performance comparison in terms of BER, 1- channel estimated from TDL-NN with MLSE equaliser, 2,3- linear equalizer with LMS and RLS algorithm for channel estimation, 4- perfect channel with MLSE equalizer, 5- CGAN channel model with AE to model Tx/Rx \cite{Ye2020}.
 % }%  This figure will be changed
%   \label{fig:comparative-study}
%    \vspace{-5mm}
%\end{center}
%\end{figure}

We used Matlab's MLSE equalizer with our channel estimates to equalize the received signal using $10{,}000$ training reference signal samples, and BER is calculated for $2000$ received signal samples for different SNR conditions for a scenario with 3 distinct multipaths. MLSE equalizer employs the Viterbi algorithm for decoding, which can also perform poorly in high noise conditions. In order to provide a fair comparison to TDL-NN, we also present BER vs.\ SNR for a perfect channel using ray-tracing ground truth in Table \ref{tab:ber}. For this simple example with 3 multipaths, TDL-NN estimates channel taps accurately and closely tracks MLSE equalization with the perfect channel, and the errors in the case of TDL-NN are largely due to MLSE equalization. Table \ref{tab:ber} also shows BER for Matlab's linear equalizer using LMS (Lsast Mean Square) and RLS (Recursive Least Square) algorithms. 

\begin{table}[!tbh]
    \begin{center}
    \captionsetup{justification=centering}
    \caption{BER versus SNR } 
    \begin{tabular}{l|c|c|c}\hline\hline
    CE Methods & SNR = 40dB & SNR = 60dB & SNR = 80dB\\
    \hline
    TDL-NN (MLSE Eq)  &  $0.51$ & ${\bf 0.44}$ & $0.15$ \\
    LMS (Linear Eq)  &  ${\bf 0.49}$ & $0.48$ & $0.39$ \\
    RLS (Linear Eq) & $0.5$ & $0.49$ & $0.44$ \\
    CGAN-AE \cite{Ye2020}  &  $0.5$ & $0.5$ & ${\bf 0.12}$ \\
    Perfect (MLSE Eq) & $0.5$& ${\bf 0.44}$ & $0.15$\\
       \hline
       \hline
    \end{tabular}
    \label{tab:ber}
    \end{center}
\end{table}
\vspace{-1mm}

ML-based CE methods are shown to provide better estimates than classical methods \cite{OShea2019,Ye2020} and we also included the BER performance of the CGAN-AE \cite{Ye2020}
approach in Table \ref{tab:ber}. CGAN-AE  models the end to end communication system with a deep autoencoder, %(transmitter with 5 and receiver with 8 hidden layers models) 
with CGAN %(generator with 4 and discriminator with 6 hidden layers) 
modelling the channel effects, so BER shows the overall performance, not just that of the CGAN. The channel is trained using $5 \times 10^6$ data samples and the results are shown here after 30 iterations, which took approximately 4 days on the same GPU, where our TDL-NN is trained with $10,000$ data samples and $5,000$ training epochs in less than $1/2$ hour. TDL-NN outperforms CGAN-AE for SNR $< 80$dB and CGAN-AE has better BER for SNR = 80dB. We remark that for more complex channels with non-linear effects, GANs CE might be better than linear TDL-NN, but its computation time and training requirements make it unsuitable for practical scenarios. In contrast, our model can be refined for real-time applications. Moreover, TDL-NN estimates multipath delays and gains and provides visibility into the channel, which is especially useful for environment sensing and awareness.

%\textbf{ToDo: Tansu based on Fig 5, our method is almost perfect? Overall question: do we have 10 hours or even 0.5 hours to estimate channels? I thought this has to be done in real time? What am I missing?}

\subsection{Sensing Example: Drone Detection}

\begin{figure}%[tbh!] 
    \subfloat[]{
        \includegraphics[width= 0.23\textwidth,scale=0.1]{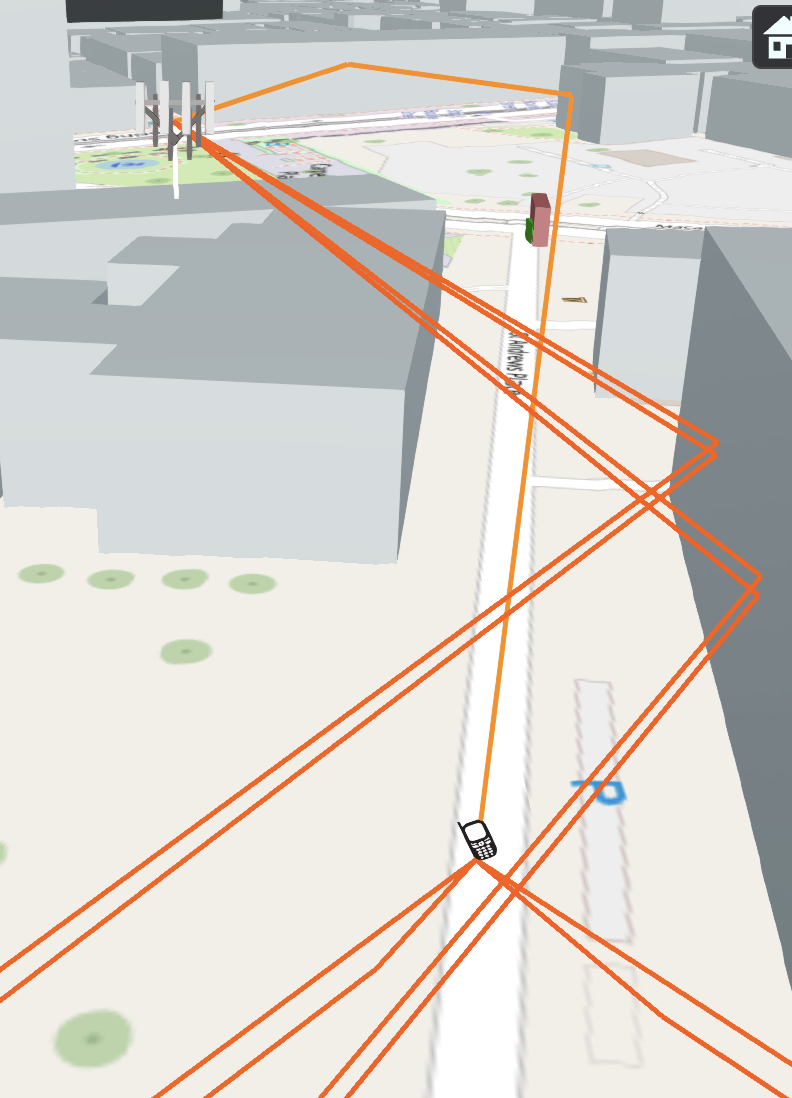}}
        \hfill
    \subfloat[]{
        \includegraphics[width= 0.23\textwidth,scale=0.1]{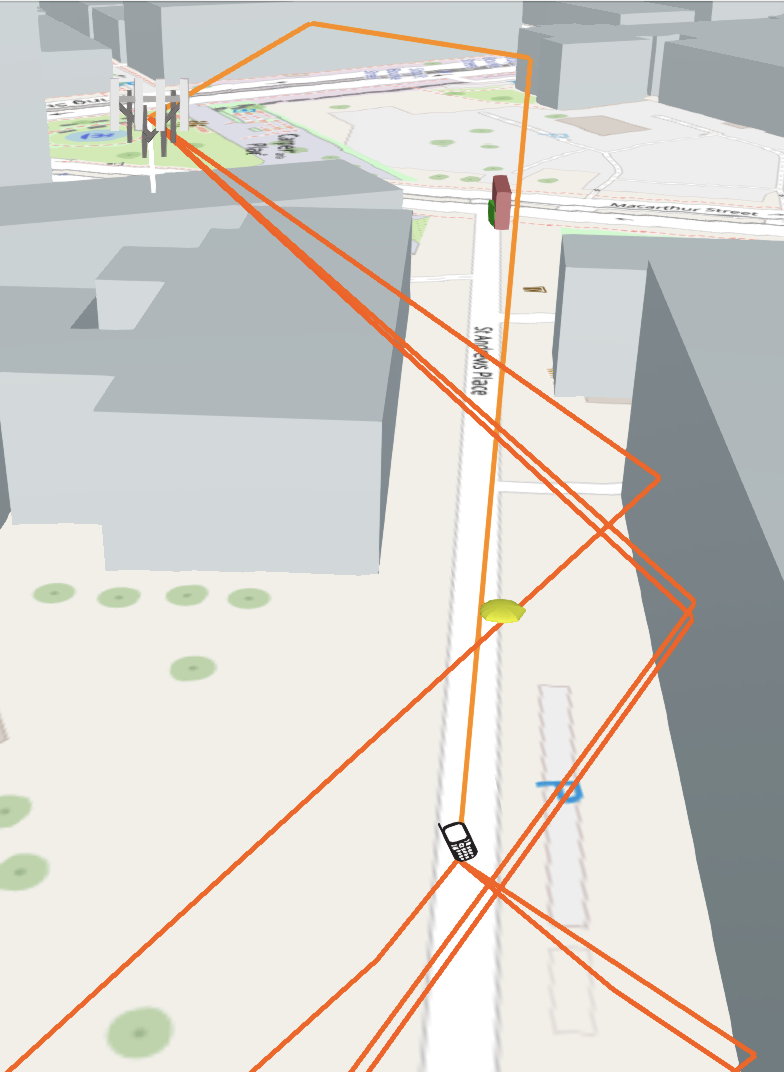}}
    \caption{(a) Scene with a mobile user transmitting to a base-station, which is continuously monitoring channel states. (b) A drone flying on the street impacts the signal multipaths.}
    \label{fig:drone_scene}
    \vspace{-5mm}
\end{figure}

\begin{figure}%[tbh!] 
    \subfloat[]{
        \includegraphics[width= 0.28\textwidth,scale=0.17]{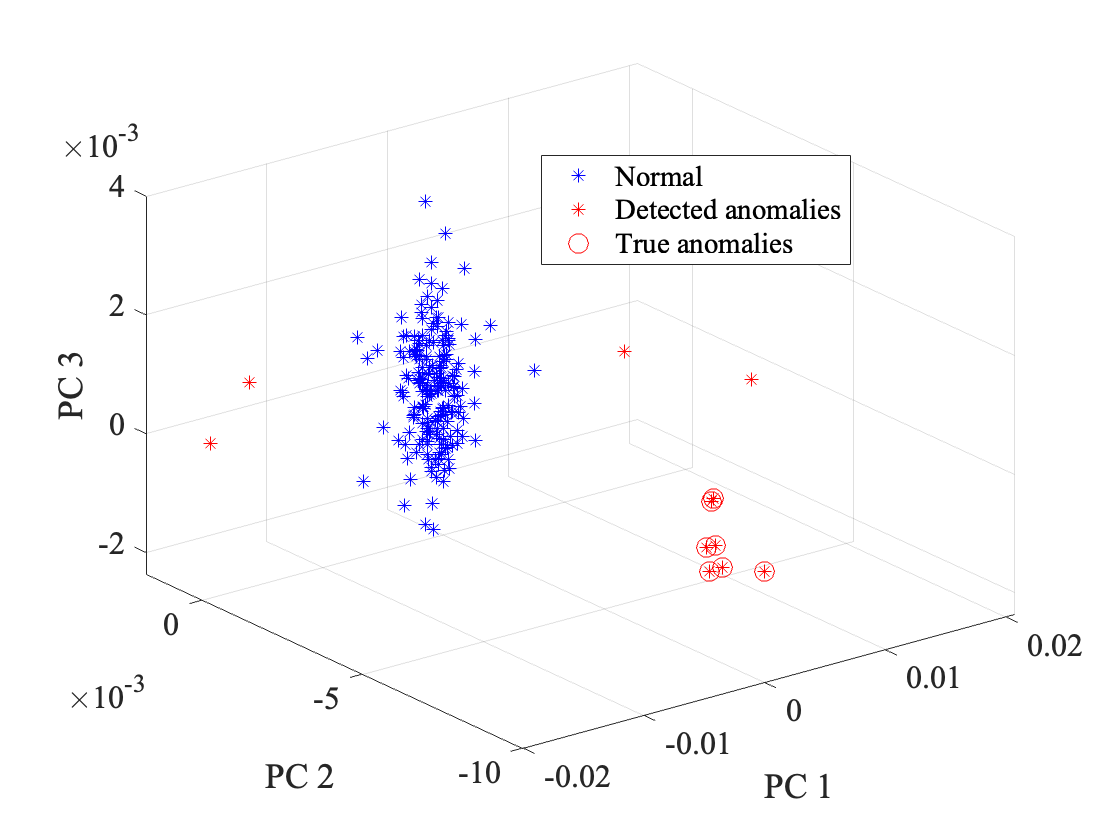}}
    \subfloat[]{
        \includegraphics[width= 0.22\textwidth,scale=0.17]{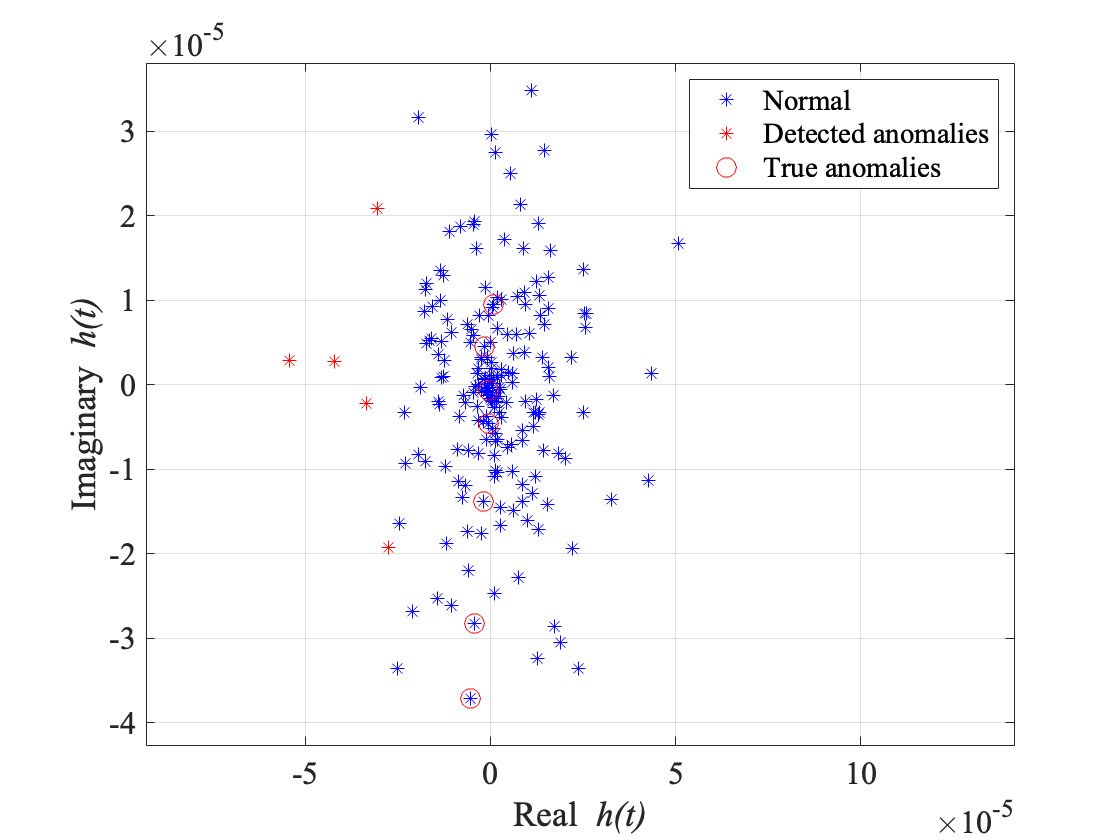}}\\
    \subfloat[]{
        \includegraphics[width= 0.28\textwidth,scale=0.17]{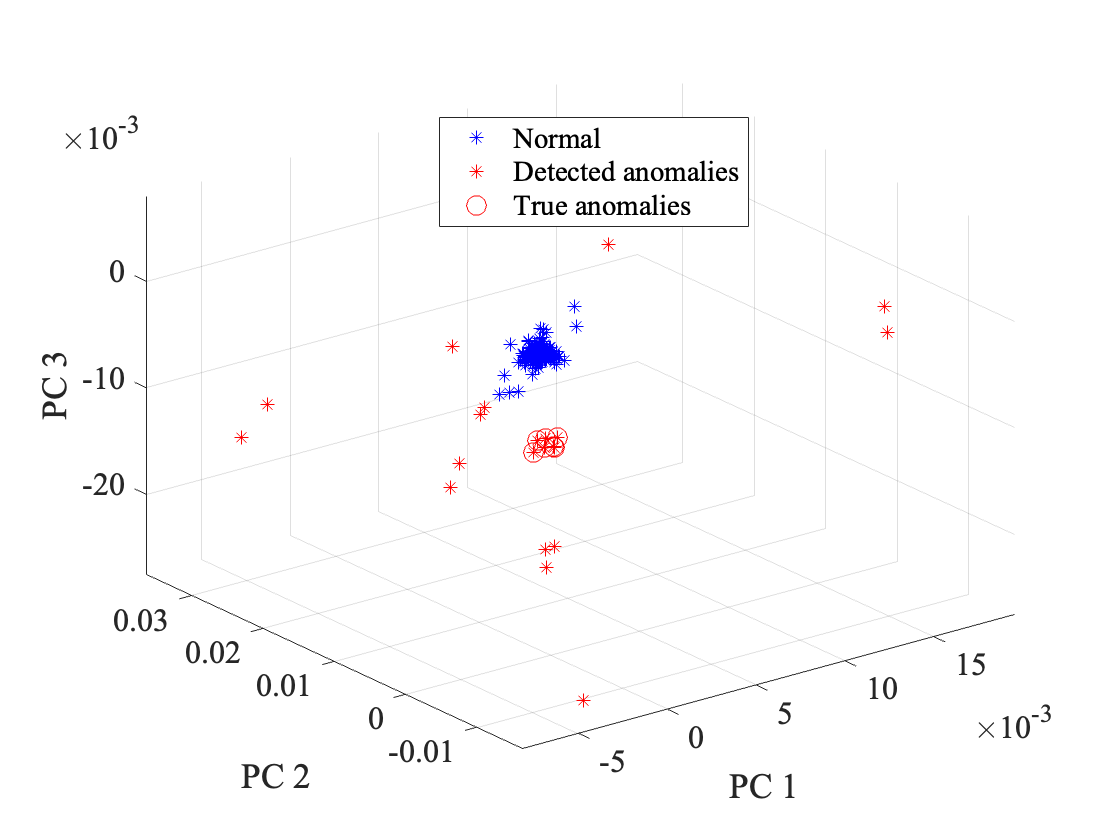}}
    \subfloat[]{
        \includegraphics[width= 0.22\textwidth,scale=0.17]{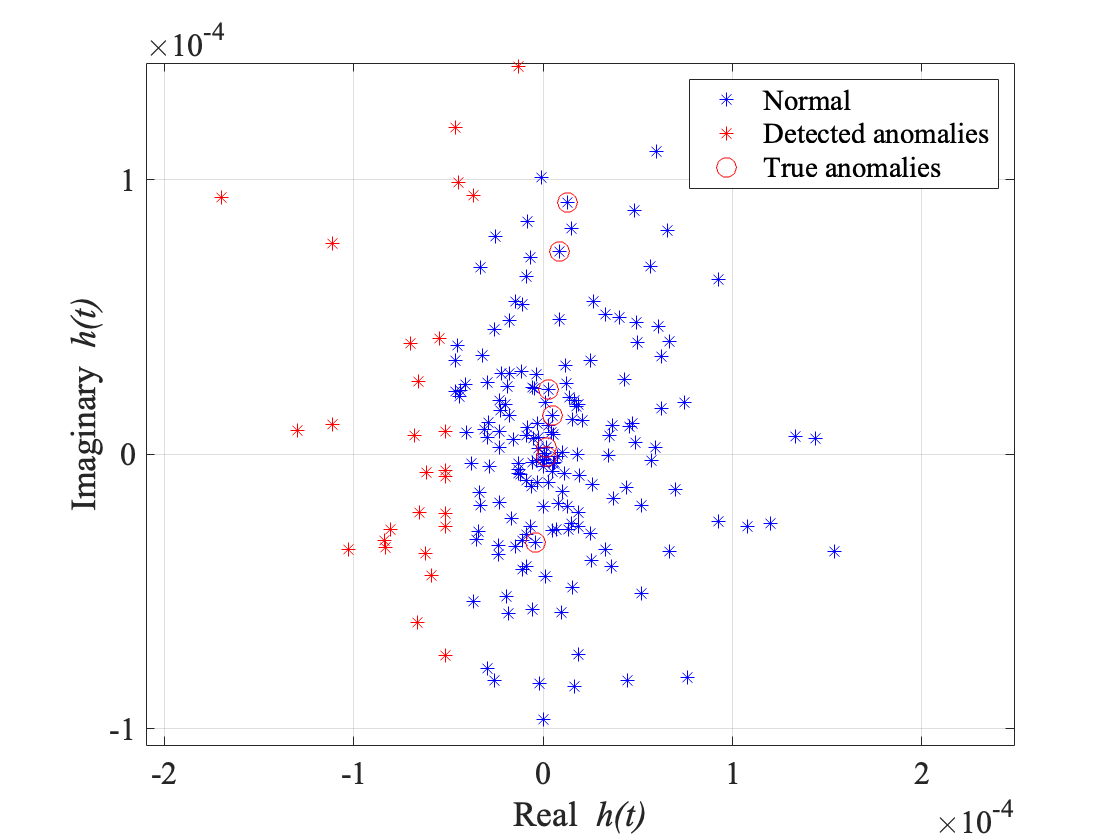}}
    \caption{Detection of anomalous channel state when drone impacts the signal. (a, b) static scenario, (a) TDL-NN based channel state and (b) conventional CSI. (c, d) dynamic scenario, (c) TDL-NN and (d) conventional CSI.}
    \label{fig:drone_results}
    \vspace{-5mm}
\end{figure}

In this section, we present a preliminary case study of detecting an unauthorised drone in an urban scenario. 
%\textbf{ToDo: Tansu: I thought this is more about usefulness/application, not interpretability. What do you mean by interpretability? Not very clear.}
Our aim here is to show the potential of our TDL-NN approach, and detection of interesting events using estimated channel state is work-in-progress. 
%The major motivation for the novel channel estimation method is to facilitate the detection of changes in the wireless signal related to interesting events or objects. For example, an unauthorised drone flying over an urban street may impact the wireless signals in the neighborhood. As a preliminary case study, 
Fig.\ \ref{fig:drone_scene} shows a scene from Matlab's site viewer with a mobile user transmitting to a roof-mounted base-station, while a drone flies over the street at an height of 10m. The communication parameters are set as in previous examples. During its flight, the drone impacts some of the signal paths as shown in Fig.\ \ref{fig:drone_scene}, which results in changes in channel state, both in delay and gain. For this preliminary example, we assume no traffic on the road. The objective is to detect the instances where a change occurs. The channel state computed by TDL-NN assumed $99$ taps, although only three taps were found to be active. Conventional CSI, on the other hand, is computed first in the frequency domain by the method of least squares using Fourier transforms of the received and transmitted signals, and then translated to the time domain via inverse Fourier transform. A simple $k$-means clustering algorithm, with $k = 2$, is used to detect anomalous channel states when the drone interfered with the signal. The results are shown in Fig.\ \ref{fig:drone_results}, where PCA (Principal Component Analysis) is used to reduce the dimensions of TDL-NN channel state to enable visualisation. It is clear from Fig.\ \ref{fig:drone_results} (a, b) that the channel state computed by TDL-NN shows the impacted states as outliers, whereas normal and anomalous states are indistinguishable when CSI is computed by the conventional method. 

The situation gets worse when we include a truck to move on the street. Its impact on the channel state makes it difficult to isolate the changes introduced by the drone. The results are shown in Fig.\ \ref{fig:drone_results} (c, d). The drone-impacted conventional CSI is indistinguishable from normal CSI, whereas the $k$-means clustering over PCA-reduced TDL-NN states resulted in many false positives along with the true detection. Thus a better algorithm is needed to isolate interesting changes from normal variations.

%We used existing change/novelty detection algorithms for this purpose which can be trained on normal data and anomalies can be detected as outliers. For this experiment, we assume to have training data from the four base-stations without the interference of drone. The results of drone detection are presented in Table \ref{tab:AD_error}. This is an on-going research project and we are building novel detection models based on the correlation between successively estimated channel states to detect sporadic events. Nevertheless, this example shows the promising potential of our approach for environment awareness application. 

\vspace{-5mm}
\section{Conclusion}\label{sec:conclusions}

In this paper, we present an efficient channel estimation approach based on a TDL model for environment sensing. In our approach, the tap delays and gains are estimated through a neural network with one hidden layer. Our extensive experiments, using Matlab's ray-tracing tool for generation of transmitted and received signal samples, show that this simple model can estimate the time-varying channel with reasonable accuracy in the presence of white noise and also in-band interference. This approach enables developing novel environmental sensing applications where interesting objects and events can be detected through the changes in the channel state. A preliminary case study is presented in the paper to show the potential of on-going research.

%Although Matlab ray-tracing is an excellent resource for this research, we understand that our experiments are limited by how accurately ray-tracing is performed by the tool. The next step in this research is to apply the CE model in a real environment. %We will not have the ground truth about the path delays and associated gains but a lower bit error rate would indicate better channel equalization with our CE model.
%We are also planning to add more layers in our model to cater for non-linearities in the 

\vspace{-3mm}
\bibliography{IChannel}
\bibliographystyle{ieeetr}

\end{document}